\documentclass[a4paper]{jpconf}
\usepackage{graphicx}

\def\rpv{R_p \hspace{-1.0em}/\;\:}
\def\321{$\mathrm{SU(3) \otimes SU(2)_L \otimes U(1)_Y}$ }
\def\e6{$\mathrm{E(6)}$ }
\newcommand{\nn}{\nonumber}

\begin{document}
\title{LHC phenomenology of supersymmetric models beyond the MSSM}

\author{Werner Porod$^{a,b}$}

\address{$^{a}$ Institut f\"ur Theoretische Physik und Astrophysik, 
Universit\"at W\"urzburg,\\
D-97074  W\"urzburg, Germany\\
$^{b}$ AHEP Group, Instituto de F\'isica Corpuscular - C.S.I.C., \\
Universitat de Val\`encia, E-46071 Val\`encia, Spain}

\ead{porod@physik.uni-wuerzburg.de}

\begin{abstract}
We discuss various phenomenological aspects of supersymmetric models
beyond the MSSM. A particular focus is on models which can correctly
explain neutrino data and the possiblities of LHC to identify the underlying
scenario.
\end{abstract}

\section{Introduction}

Supersymmetric extensions of the standard model (SM) are promising
candidates for new physics at the TeV scale \cite{Nilles:1983ge,Haber:1984rc} 
as the solve several
short-comings of the Standard Model (SM). The Minimal Supersymmetric
Standard Model (MSSM) solves the hierarchy problem of the SM
\cite{Witten:1981nf}, leads to a unification of the gauge
couplings \cite{Dimopoulos:1981yj,Ibanez:1981yh} and introduces several candidates
for dark matter depending on how SUSY is broken
\cite{Ellis:1983ew,Steffen:2008qp}. Its phenomenology with respect to
present and future colliders has been
widely explored, see e.g.~\cite{Weiglein:2004hn,Nath:2010zj}.

However, similarly as the SM it needs additional ingredients to explain
neutrino data, e.g.~by either incooperating heavy new particles giving
rise to tiny neutrino masses via the seesaw mechanism \cite{Minkowski:1977sc} or
via the braking of R-parity \cite{Hall:1984id}. Moreover,  a new
problem arises in the MSSM not present in the SM: the superpotential contains a parameter
with dimension mass, namely the so called $\mu$ parameter which gives masses to
the Higgs bosons and higgsinos. From a purely theoretical point of view, the
value of this parameter is expected to be either of the order of the GUT/Planck scale or
exactly zero, if it is protected by a symmetry. For phenomenological
aspects, however, it has to be  of the
order of the scale of electroweak symmetry breaking (EWSB) and it has to
be non-zero to be consistent with experimental data. This
discrepancy is the so called \(\mu\)-problem of the MSSM
\cite{Kim:1983dt}.

In this paper we discuss various supersymmetric models addressing at least
one of these two topics focusing on features of their phenomenology which can
be tested at the LHC and which differ from the usual MSSM phenomenology. 
We will take as main guideline the requirement that neutrino data are correctly
explained.
We will first discuss briefly the case of Dirac neutrinos
and afterwards discuss models leading to Majorana neutrinos, both in the
context of conserved R-parity. In the third part we cover models with broken
R-parity and their phenomenology at the LHC. 

\section{Dirac neutrinos}

Technically the easiest way to obtain masses for neutrinos is by introducing
a new Yukawa coupling similar to the ones for the SM fermions. In this case
the MSSM has to be extended by additional right handed neutrinos, which
are gauge singlets with respect to the SM gauge group, and the superpotential
of the MSSM
\begin{equation}
W_{MSSM} = {\hat H}_d {\hat L} Y_{e} {\hat E}^c 
         + {\hat H}_d {\hat Q} Y_{d} {\hat D}^c 
         + {\hat H}_u {\hat Q} Y_{u} {\hat U}^c  \nonumber
         - \mu {\hat H}_d {\hat H}_u
\end{equation}
has to be extended by the term
\begin{equation}
W_{\nu^c} =     {\hat H}_u {\hat L} Y_{\nu} {\hat \nu}^c  \,.
\end{equation}
Here ${\hat H}_d$, ${\hat H}_u$, $ {\hat Q}$, ${\hat D}^c$, ${\hat
U}^c$, ${\hat L}$ and ${\hat E}^c$ are the MSSM superfields containing
the Higgs bosons, quarks, leptons and their supersymmetric
partners. In addition the superfield ${\hat \nu}^c$ for the
right-handed neutrino and the right-sneutrino has been introduced.
The corresponding Yukawa coupling $Y_{\nu}^{ij}$ has to be tiny, of
the order $10^{-12}$ and smaller, to explain correctly neutrino masses
in the sub-eV range as required by data
\cite{Schwetz:2008er}. 
This in turn implies that one is essentially left with the usual
MSSM phenomenology, except for the case where the right-sneutrino is
the lightest supersymmetric particle (LSP): in this case all decays
are as in the MSSM down to the next to lightest supersymmetric
particle (NLSP) which eventually decays into the LSP. The
corresponding width of the last step is proportional to $|Y_\nu|^2$
and thus rather small implying at the LHC decay lengths ranging from
$O(cm)$ up to $O(km)$ \cite{deGouvea:2006wd}. Consequently the signal
of such a scenario is very two long-lived particles in each event
which can even appear as stable particles in a typical detector of
high energy collider experiments. Detailed studies have been performed
for the cases of a stau \cite{Gupta:2007ui} NLSP and a stop
\cite{Choudhury:2008gb} NLSP demonstrating that LHC should be able to
identity such scenarios.

\section{Majorana neutrinos via the seesaw mechanism}

A possibility to obtain tiny neutrino masses while having at the same time
sizable Yukawa couplings are seesaw scenarios  where very heavy new particles
are postulated inducing the so-called Weinberg operator 
\cite{Weinberg:1979sa,Weinberg:1980bf}
\begin{equation}\label{eq:dim5}
 \frac{f_{\alpha\beta}}{\Lambda} (H_u L_\alpha) (H_u L_\beta)
 \Rightarrow  (m^{\nu})_{\alpha\beta} = \frac{f_{\alpha\beta} v^2_u}{2 \Lambda}
\end{equation}
when integrating out the heavy degrees of freedom. Here $\Lambda$ is a measure of the
scale for these new particles and $f_{\alpha\beta}$ is usually a combination of 
different Yukawa couplings. One can show, that
at tree-level only three possibilities exist to realize such scenarios
\cite{Ma:1998dn}. Type-I is the well-known case of the exchange of 
a heavy fermionic singlet usually denoted $\nu_R$
\cite{Minkowski:1977sc,yanagida:1979,gell-mann:1980vs,mohapatra:1979ia}. 
Type-II corresponds to the exchange of a scalar $SU(2)$ triplet 
\cite{Schechter:1980gr,Cheng:1980qt}. In seesaw type-III 
one adds (at least two) fermionic $SU(2)$ triplets to the field content of 
the SM \cite{Foot:1988aq}.  
In the last two models particles charged under the SM gauge
group are added and which correspond to incomplete $SU(5)$ representations.
As they would destroy 
the nice feature of gauge coupling unification  one often adds at the
seesaw scale(s) additional particles to obtain complete $SU(5)$ representations
to maintain the feature of gauge coupling unification.
 A  detailed discussion including the embedding in $SU(5)$
models can be found in e.g.~in  ref.\ \cite{Borzumati:2009hu}. 

Below
we present only the various superpotential for briefness. 
In addition there will also be
the corresponding soft SUSY terms which, however, reduce at the electroweak
scale to the MSSM ones once the states with masses at the seesaw scale(s)
are integrated out and, thus, are not discussed further. 
These additional
terms only contribute to threshold corrections at the seesaw scale(s) and their
effect is negligible if one requires universal boundary conditions for all soft terms
\cite{Kang:2010zd}.
Here we will assume common soft SUSY breaking at the GUT-scale $M_G$
to specify the spectrum at the electroweak scale: a common gaugino
mass $M_{1/2}$, a common scalar mass $m_0$ and the trilinear coupling
$A_0$ which gets multiplied by the corresponding Yukawa couplings to
obtain the trilinear couplings in the soft SUSY breaking
Lagrangian. In addition the sign of the $\mu$ parameter is fixed as
well as $\tan\beta =v_u/v_d$ at the electroweak scale where $v_d$ and
$v_u$ are the the vacuum expectation values (vevs) of the neutral
component of $H_d$ and $H_u$, respectively.

\subsection{Supersymmetric seesaw type-I}
\label{sec:modelI}

In this class of models one postulates very heavy right-handed neutrinos
yielding the following superpotential below the GUT-scale:
\begin{eqnarray}
W_{I} &=& W_{MSSM} + W_{\nu^c} + \frac{1}{2}{\hat \nu}^c M_R  {\hat \nu}^c
 \label{eq:superpotI} 
\end{eqnarray}
For the neutrino mass matrix one obtains the well-known formula
\begin{equation}
m_\nu = - \frac{v^2_u}{2} Y^T_\nu M^{-1}_R Y_\nu.
\label{eq:mnuI}
\end{equation}
Being complex symmetric, the light Majorana neutrino mass matrix
  in eq.~(\ref{eq:mnuI}) is diagonalized by a unitary $3\times 3$ matrix
  $U$~\cite{Schechter:1980gr}
\begin{equation}\label{diagmeff}
{\hat m_{\nu}} = U^T \cdot m_{\nu} \cdot U\ .
\end{equation}
Inverting the seesaw equation eq.~(\ref{eq:mnuI}) allows to express 
$Y_{\nu}$ as \cite{Casas:2001sr}
\begin{equation}\label{Ynu}
Y_{\nu} =\sqrt{2}\frac{i}{v_u}\sqrt{\hat M_R}\cdot R \cdot \sqrt{{\hat
    m_{\nu}}} \cdot U^{\dagger},
\end{equation}
where the $\hat m_{\nu}$ and $\hat M_R$ are  diagonal matrices containing the
corresponding eigenvalues. 
$R$ is  in general a complex orthogonal matrix. We assume
 $R=1$ implying that $Y_{\nu}$ contains only ``diagonal'' products
$\sqrt{M_im_{i}}$.

\subsection{Supersymmetric seesaw type-II}
\label{sec:modelII}

In seesaw models of type II one adds a scalar $SU(2)$ triplet $T$ to generate
neutrino masses. As this triplet carries also hypercharge one has to embed
it in a $15$-plet of $SU(5)$ which has under
 $SU(3)\times SU_L(2) \times U(1)_Y$ the following decomposition \cite{Rossi:2002zb}
\begin{eqnarray}\label{eq:15}
{\bf 15} & = &  S + T + Z \\ \nonumber
S & \sim  & (6,1,-\frac{2}{3}), \hskip10mm
T \sim (1,3,1), \hskip10mm
Z \sim (3,2,\frac{1}{6}).
\end{eqnarray}
In supersymmetric models one  adds a pair  $15$ and $\overline{15}$ to avoid
anomalies below the GUT-scale.
Below the GUT scale in the $SU(5)$-broken 
phase the superpotential reads
\begin{eqnarray}\label{eq:broken}
W_{II} & = & W_{MSSM} + \frac{1}{\sqrt{2}}(Y_T \widehat{L} \widehat{T}_1  \widehat{L} 
+  Y_S \widehat{D}^c \widehat{S}_1 \widehat{D}^c) 
+ Y_Z \widehat{D}^c \widehat{Z}_1 \widehat{L}   \nonumber \\
& + & \frac{1}{\sqrt{2}}(\lambda_1 {\widehat H}_d \widehat{T}_1  {\widehat H}_d 
+\lambda_2 {\widehat H}_u \widehat{T}_2 {\widehat H}_u) 
+ M_T \widehat{T}_1 \widehat{T}_2 
+ M_Z \widehat{Z}_1 \widehat{Z}_2 + M_S \widehat{S}_1 \widehat{S}_2
\end{eqnarray}
where the states with index 1 (2) belong the $15$-plet ($\overline{15}$-plet).
The second term in eq.~(\ref{eq:broken}) is responsible for the generation 
of the neutrino masses yielding 
\begin{eqnarray}\label{eq:ssII}
m_\nu=-\frac{v_2^2}{2} \frac{\lambda_2}{M_T}Y_T.
\end{eqnarray}
Note that 
\begin{equation}\label{diagYT}
{\hat Y}_T = U^T \cdot Y_T \cdot U
\end{equation}
i.e. $Y_T$ is diagonalized by the same matrix as $m_{\nu}$.

In addition there are the couplings $Y_S$ and $Y_Z$, which in  
principle  are not determined by any low-energy data. In the calculation 
of lepton flavour violating 
observables both Yukawa couplings, $Y_T$ and 
$Y_Z$, contribute. Having a GUT model in mind we require for
the numerical discussion later the $SU(5)$ boundary conditions
 $Y_T=Y_S=Y_Z$ at the GUT scale.  
As long as $M_Z \sim M_S \sim M_T \sim M_{15}$ gauge coupling unification 
will be maintained. The equality need not be exact for a successful unification. 
In our numerical studies we have taken into account the different running
of these mass parameters but we decouple them all at the scale $M_T(M_T)$
because the differences are small.

\subsection{Supersymmetric seesaw type-III}
\label{sec:modelIII}

In the case of a seesaw model type III one needs new fermions $\Sigma$ 
at the high
scale being in the adjoint representation of $SU(2)$. They have to be 
embedded in a $24$-plet  to obtain a complete $SU(5)$  representation
which decomposes under  $SU(3)\times SU_L(2) \times U_Y(1)$ as 
\begin{eqnarray}\label{eq:def24}
24_M & = &(1,1,0) + (8,1,0) + (1,3,0) + (3,2,-5/6) + (3^*,2,5/6) \\ \nn
   & = & \widehat{B}_M + \widehat{G}_M + \widehat{W}_M + \widehat{X}_M + \widehat{\bar X}_M 
\end{eqnarray}
The fermionic components of $(1,1,0)$ and $(1,3,0)$ have exactly 
the same quantum numbers as $\nu^c$ and $\Sigma$. Thus, the $24_M$ 
always produces a combination of the type-I and type-III seesaw. 
In the $SU(5)$ broken phase the superpotential reads
\begin{eqnarray}\label{eq:spotIII}
 W_{III} & = &  W_{MSSM}
 +  \widehat{H}_u(\sqrt{2} \widehat{W}_M Y_N - \sqrt{\frac{3}{10}} 
               \widehat{B}_M Y_B) \widehat{L}
 + \widehat{H}_u \widehat{\bar X}_M Y_X \widehat{D}^c \nonumber \\
         & & + \frac{1}{2} \widehat{B}_M M_{B} \widehat{B}_M 
         + \frac{1}{2}\widehat{G}_M M_{G} \widehat{G}_M 
          + \frac{1}{2} \widehat{W}_M M_{W} \widehat{W}_M 
          + \widehat{X}_M M_{X} \widehat{\bar X}_M 
\end{eqnarray}
As above we use at the GUT scale the boundary condition
$Y_N = Y_B = Y_X$ and $M_B = M_G=M_W=M_X$. Integrating out the heavy fields
yields the following formula for the neutrino masses at the low scale:
\begin{equation}
m_\nu = - v^2_u \left( \frac{3}{10} Y^T_B M^{-1}_B Y_B + \frac{1}{2} Y^T_W M^{-1}_W Y_W \right). 
\label{eq:mnu_seesawIII}
\end{equation}
The boundary conditions at $M_G$ imply that at the seesaw scale(s) one still has
$M_B \simeq M_W$ and $Y_B \simeq Y_W$ so that one can write in a good approximation
\begin{equation}
m_\nu = - v^2_u  \frac{4}{5} Y^T_W M^{-1}_W Y_W 
\label{eq:mnu_seesawIIIa}
\end{equation}
and one can use the corresponding  decomposition for $Y_W$ as discussed in section
\ref{sec:modelI}.

\subsection{Effect of the heavy particles on the MSSM spectrum}

\begin{figure}[t]
  \centering
\includegraphics[width=0.45\textwidth]{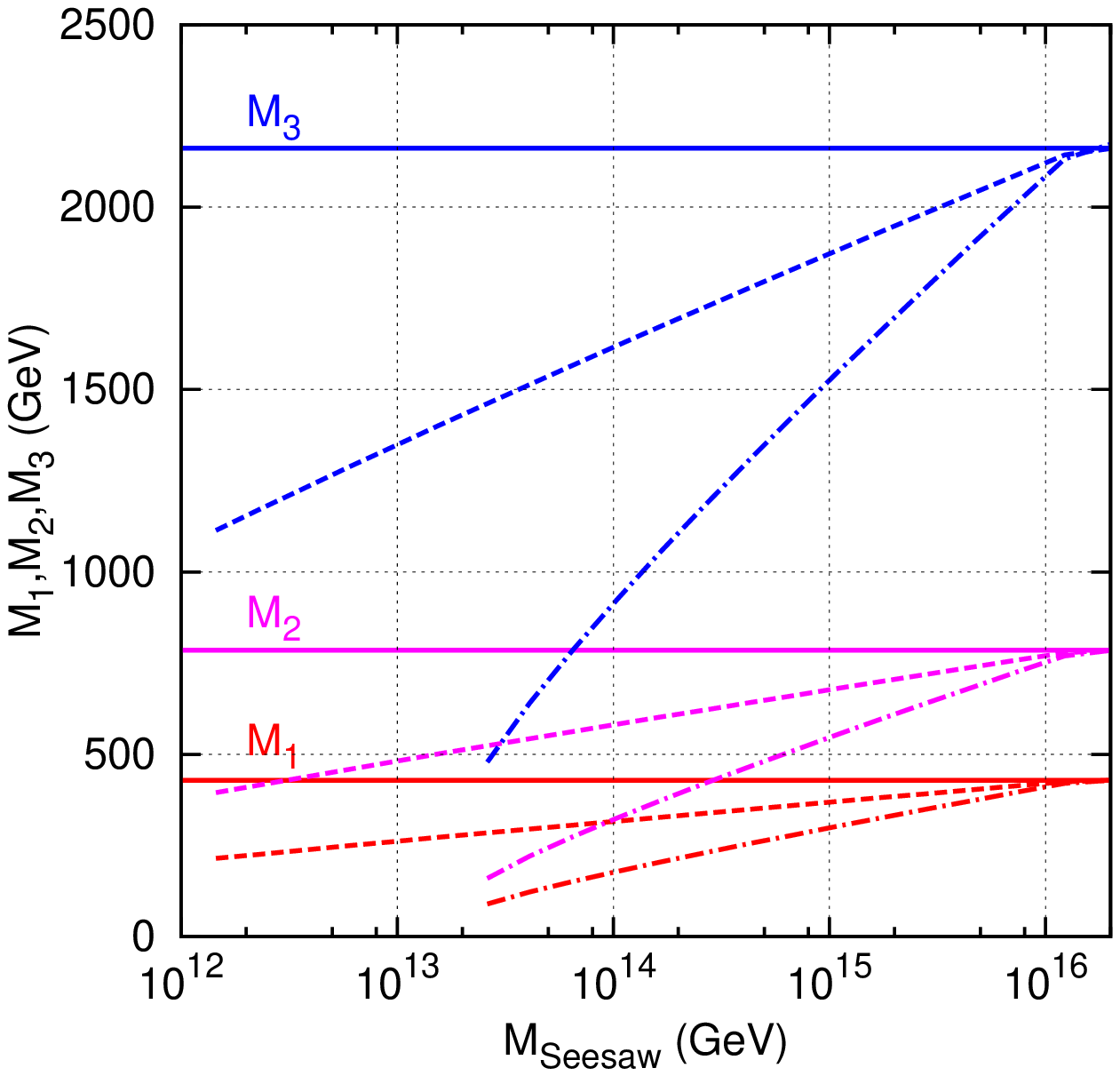}
\hspace{0.08\textwidth}
\includegraphics[width=0.45\textwidth]{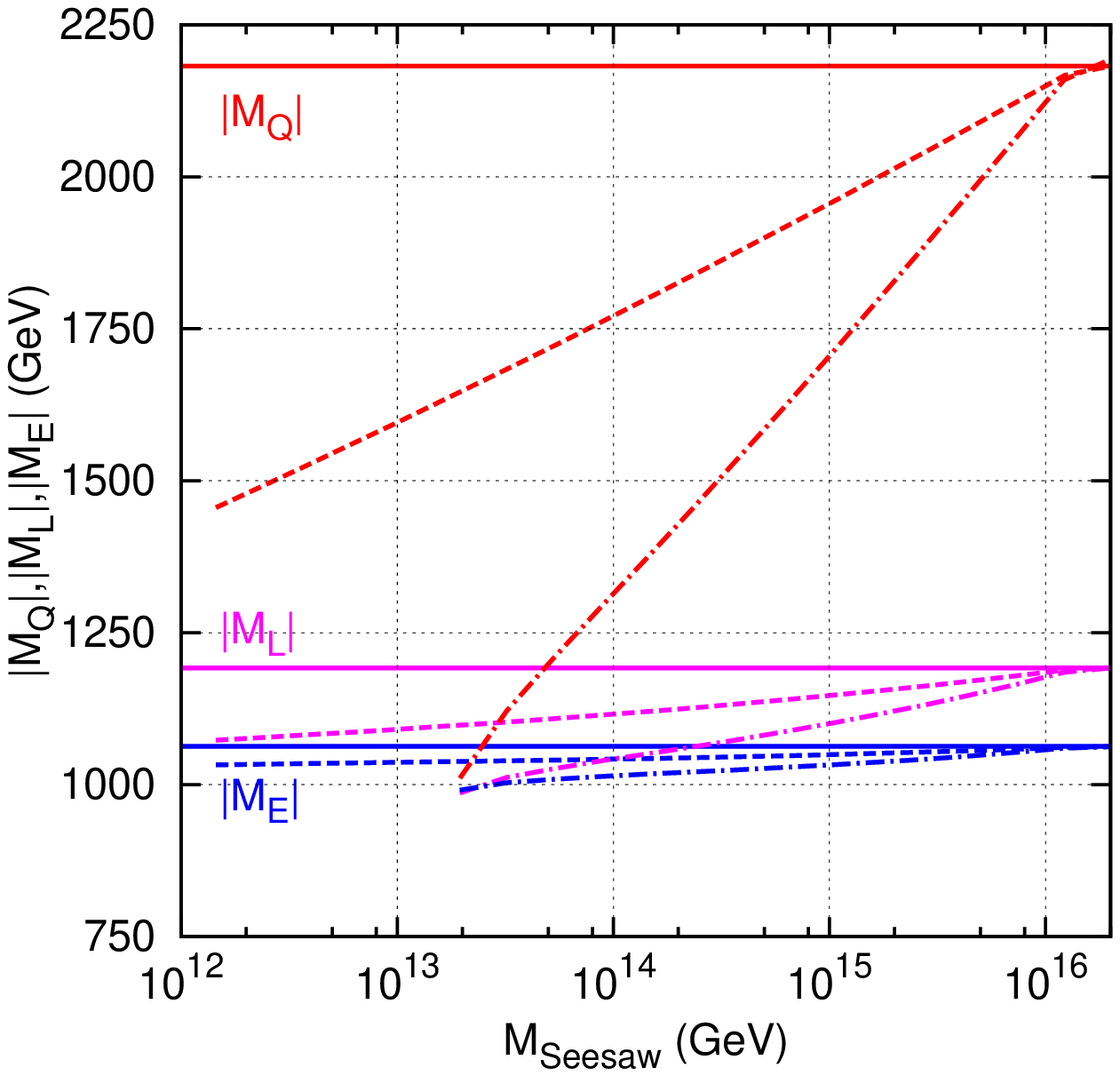}
\caption{Mass parameters at $Q=1$ TeV versus the seesaw scale for fixed
high scale parameters $m_0= M_{1/2} = 1$ TeV, $A_0=0$, $\tan\beta=10$ and $\mu>0$. 
The full lines correspond to seesaw type I, the dashed ones to type II and the
dash-dotted ones to type III. In all cases a degenerate spectrum of the seesaw
particles has been assumed.}
  \label{fig:mass1000}
\end{figure}

The appearance of charged particles at scales between the electroweak
scale and the GUT scale leads to changes in the beta functions
of the gauge couplings \cite{Rossi:2002zb,Buckley:2006nv}.
In the MSSM the corresponding values at 1-loop level are
$(b_1,b_2,b_3)=(33/5,1,-3)$. In case of one $15$-plet the additional
contribution is $\Delta b_i=7/2$ whereas in case of $24$-plet
it is $\Delta b_i=5$. This  results in case of type II in 
a total shift of  $\Delta b_i=7$ for the minimal model and
in case of type III in  $\Delta b_i=15$  assuming 3 generations
of $24$-plets. This does not only change the evolution
of the gauge couplings but also the evolution of the gaugino
and scalar mass parameters with profound implications on
the spectrum \cite{Buckley:2006nv,Hirsch:2008gh}. Additional
effects on the spectrum of the scalars can be present if some of the Yukawa 
couplings get large
\cite{Hirsch:2008gh,Calibbi:2009wk,Biggio:2010me}.
In figure \ref{fig:mass1000} we exemplify this by showing the values
of selected mass parameters at $Q=1$ TeV versus the seesaw scale for
fixed high scale parameters $m_0= M_{1/2} = 1$ TeV, $\tan\beta=10$ and $A_0=0$
 where we have assumed that
the
additional Yukawa couplings are small. As expected, the effects in case
of models of type II and III are larger the smaller the corresponding
seesaw-scale is. The scalar mass parameters shown are of the first 
generation and, thus, the results are nearly independent of $\tan\beta$
and $A_0$.

\begin{figure}[h]
\begin{minipage}{0.47\textwidth}
\includegraphics[height=60mm,width=70mm]{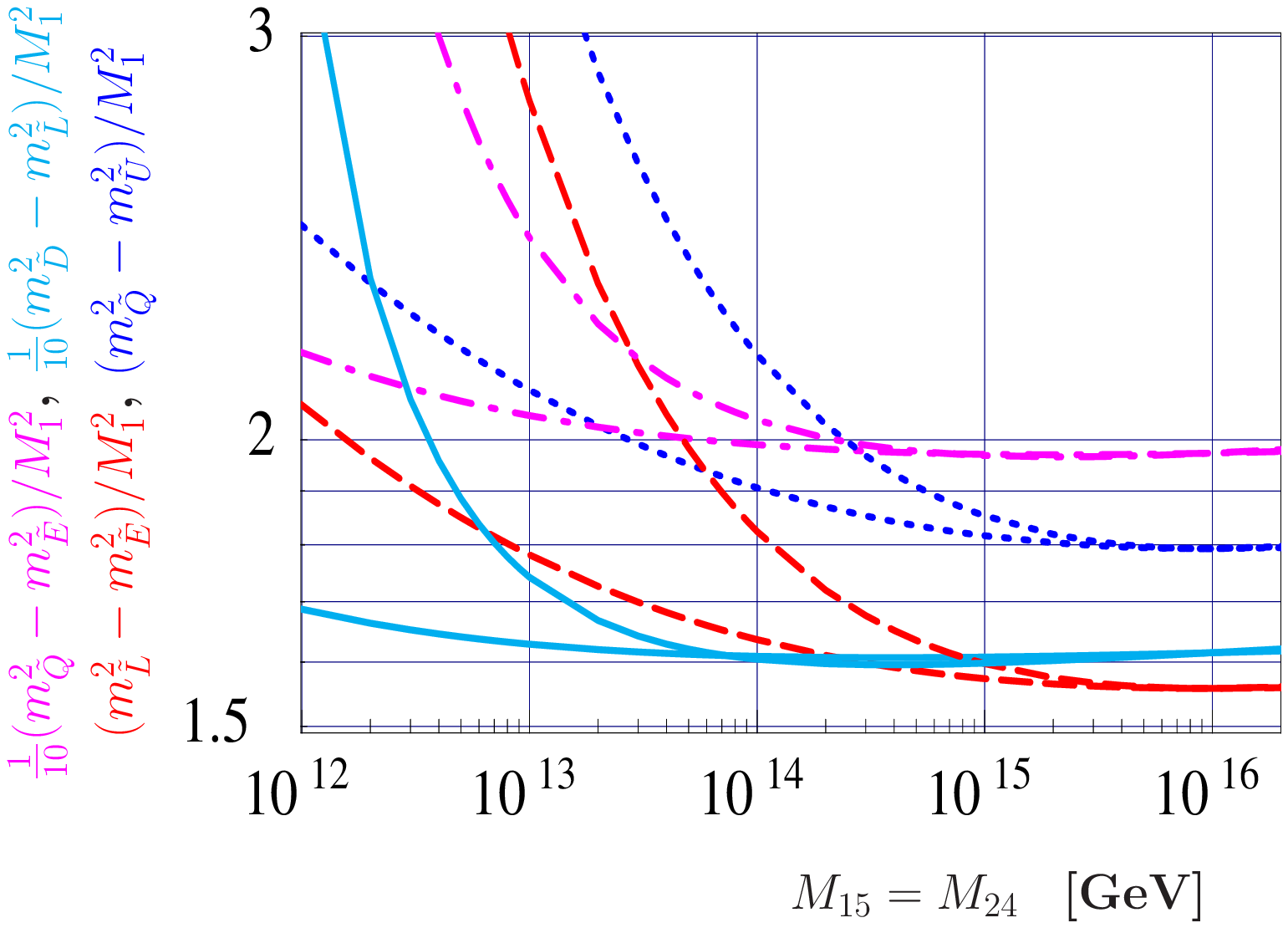}
\vskip0mm
\caption{Four different ``invariant'' combinations of soft masses 
versus the mass of the ${\bf 15}$-plet {\em or} ${\bf 24}$-plet, 
$M_{15}=M_{24}$. The calculation is at 1-loop order in the leading-log 
approximation. The lines running faster up towards smaller $M$ are 
for type-III seesaw, the lower ones are for type-II seesaw.}
\label{fig:ana}
\end{minipage}\hspace{0.04\textwidth}%
\begin{minipage}{0.47\textwidth}
\includegraphics[height=75mm,width=80mm]{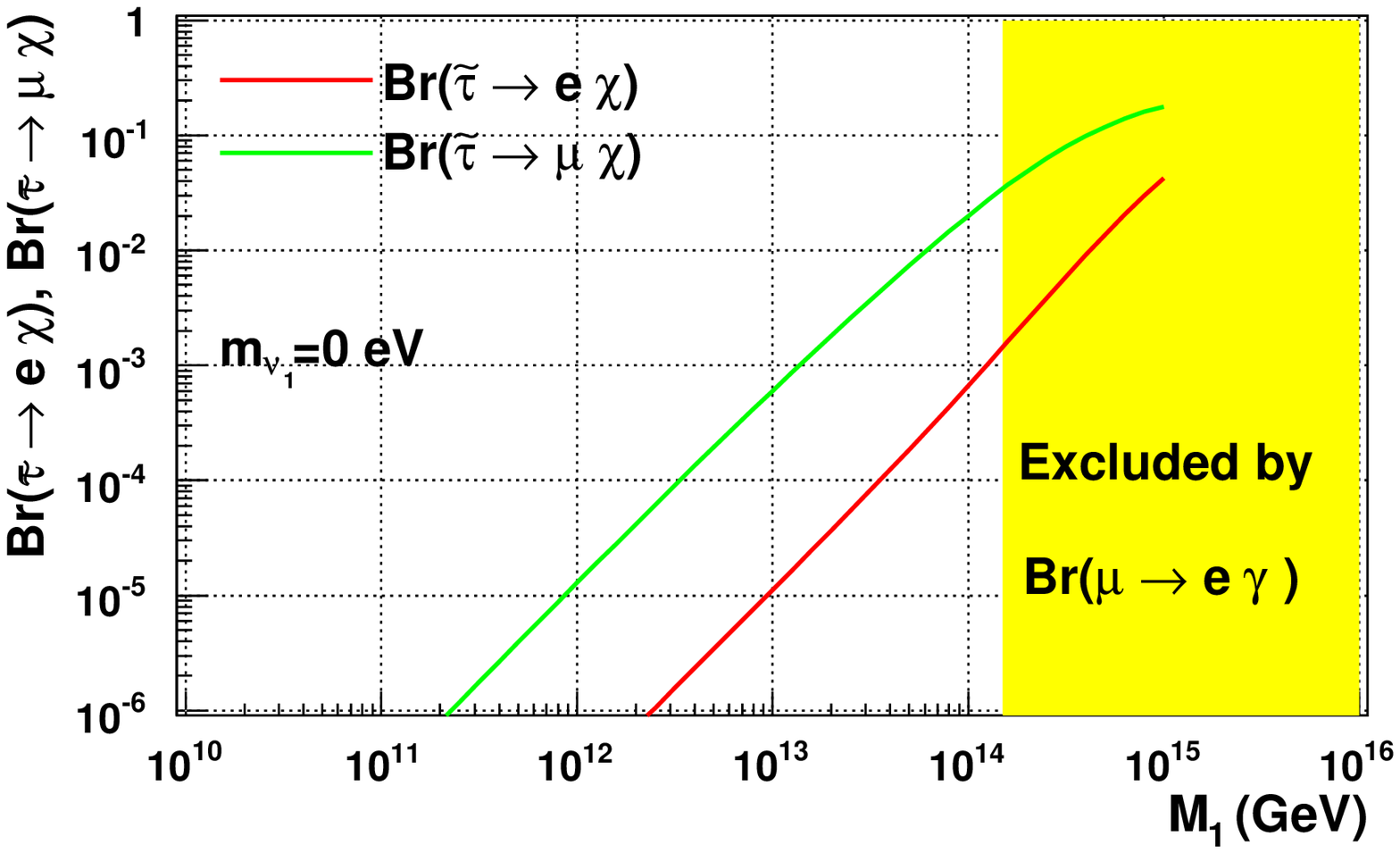}
\caption{\label{fig:seesawIb} Flavour violating decays of the heavier stau for degenerate
right-handed neutrinos, $M_0=90$~GeV, $M_{1/2} = 400$~GeV,
$A_0=0$~GeV, $\tan\beta=10$, $\mu >0$), from ref.\ \cite{Hirsch:2008dy}}
\end{minipage} 
\end{figure}

Note that in all three model types the ratio of the gaugino mass
parameters is nearly the same as in the usual mSUGRA scenarios but the
ratios and differences of the sfermion mass parameters change
\cite{Buckley:2006nv,Hirsch:2008gh}. One can form four 'invariants' where at
least at the 1-loop level the dependence on $M_{1/2}$ and $m_0$ is
rather weak, e.g.\ $(m^2_{\tilde L} - m^2_{\tilde E})/M^2_1$,
$(m^2_{\tilde Q} - m^2_{\tilde E})/M^2_1$, $(m^2_{\tilde D} -
m^2_{\tilde L})/M^2_1$ and $(m^2_{\tilde Q} - m^2_{\tilde U})/M^2_1$.
Here one could replace $M_1$ by any of the other two gaugino masses
which simply would amount in an overall rescaling.
In figure \ref{fig:ana} we show these 'invariants' in the leading-log
approximation at 1-loop order to demonstrate the principal behaviour
for seesaw type II with a pair of ${\bf 15}$-plets and seesaw type III
with three ${\bf 24}$-plets. Due to the larger change in the
beta-coefficient the effect is more pronounced in case of type III
models. From this one concludes that in principle one has a handle to
get information on the seesaw scale for given assumptions on the
underlying neutrino mass model and assuming universal boundary
conditions at the GUT scale.  For the type-I, i.e.\ singlets only, of
course $\Delta b_i=0$ and no change with respect to mSUGRA are
expected. A detailed discussion can be found e.g.\ in
\cite{Hirsch:2008gh}.

\subsection{Lepton flavour violation in the slepton sector and signals at the LHC}
\label{sec:model_LFV}

Additional effects beside the ones on the spectrum discussed above can 
occur due to lepton flavour mixing entries in the sleptons mass matrix
which are induced due to RGE effects by the additional Yukawa couplings. 
From a one-step integration of the RGEs one gets, assuming mSUGRA boundary
conditions, a first rough
estimate for the lepton flavour violating entries in the slepton
mass parameters:
\begin{eqnarray}
m^2_{\tilde L,ij} &\simeq& -\frac{a_k}{8 \pi^2 } 
 \left( 3 m^2_0 +  A^2_0 \right) 
 \left(Y^{k,\dagger} L Y^{k}\right)_{ij} \\
A_{l,ij} &\simeq& - \frac{3 a_k}{ 16 \pi^2 }   A_0 
 \left(Y_E Y^{k,\dagger} L Y^{k}_N\right)_{ij}
\end{eqnarray}
for $i\ne j$ in the basis where $Y_E$ is diagonal and 
$L_{ij} = \log(M_G/M_i)\delta_{ij}$.
The corresponding coefficient $a_k$ for seesaw type $k$ are: $a_I=1$,
$a_{II} = 6$ and $a_{III}=9/5$.   
All models have in common that the predict negligible flavour violation for the $R$-sleptons
$m^2_{\tilde E,ij} \simeq 0$. Sizable entries for the these parameters would be a 
clear hint for a left-right symmetric extension of the MSSM \cite{Vicente:2010wa}.

\begin{figure}[t]
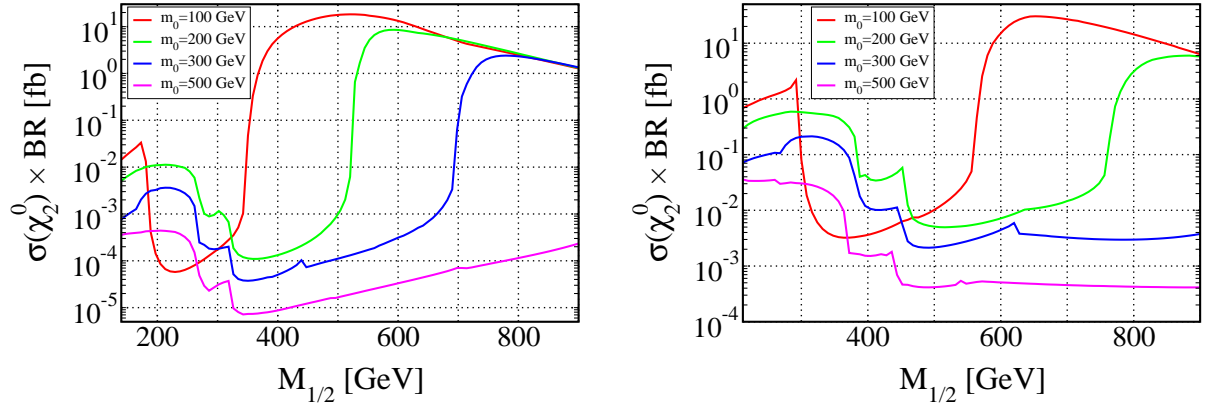

\includegraphics[width=0.47\textwidth]{figs/plot-sigmaLOfbBR-m12_-_4m0.eps}
\hspace{4mm}
\includegraphics[width=0.47\textwidth]{figs/plot-sigmaLOfbBR-m12_-_4m0_-_II.eps}
\caption{\label{fig:seesawIc} Cross section $\sigma(p p \to \tilde \chi^0_2) 
\times BR(\chi^0_2 \to \sum_{i,j} \tilde l_i l_j 
    \to \mu^\pm \tau^\mp \tilde \chi^0_1)$ as a function of $M_{1/2}$ for various
values of $m_0$,
$A_0=0$, $\tan\beta=10$, $\mu >0$. Left side, seesaw type I with degenerate
$\nu_R$, right side seesaw type II with $\lambda_1=0.02$, $\lambda_2=0.5$;
from ref.\ \cite{Esteves:2009vg}.}
\end{figure}

These flavour mixing entries induce on the one hand rare lepton decays, e.g.\
$\mu \to e \gamma$, and on the other hand lepton flavour violating decays
of sleptons and neutralinos. As a typical example we show in fig.~\ref{fig:seesawIb}
the flavour violating decays of the heavier stau in a seesaw I scenario. One sees
that branching ratios of at most a few per-cent can be reached close to the region
excluded by $\mu\to e\gamma$ where $Y_\nu$ gets sizable. Also in case of seesaw type
II one obtains similar results \cite{Hirsch:2008gh}. At the LHC however one will not
be able to identify the individual branching ratios but one has to consider the
complete cascade starting from the production and taking into account all steps
of the various decays, e.g.\ $\tilde q_L \to q \tilde \chi^0_2 \to q l_i^{\pm}
{\tilde l}^{\mp}_j \to q l_i^{\pm} l_k^{\mp}\tilde \chi^0_1$. As typical examples we
display in fig.~\ref{fig:seesawIc} the obtainable cross section containing a
$\mu^\pm \tau^\mp$ pair in the final state in case of
 seesaw models of type I and II for various values of $m_0$
as a function of $M_{1/2}$. Here the signal has been maximized by choosing the
Yukawa couplings such that $BR(\mu\to e \gamma) = 10^{-12}$ \cite{Esteves:2009vg}.
One sees that one gets at most signals up to $O(20-30)$ fb
which clearly requires a large luminosity to be detected.

\section{R-parity violation}

We now turn to the  possibility that low-energy supersymmetry
itself may provide the origin of neutrino
mass~\cite{Ross:1984yg,Hall:1984id},
for a review see Ref.~\cite{Hirsch:2004he}.  Usually one
assumes that R-parity, defined as $(-1)^{3B+L+2S}$, is an exact
symmetry under which all superpartners are odd and SM particles even.
However the terms that break R-parity are allowed by supersymmetry as
well as the SM gauge invariance. Expressed as superfields, they
have the form $\hat L \hat H_u$, $\hat L \hat L \hat E^c$, 
$\hat Q \hat L \hat D^c$ and $\hat U^c \hat D^c \hat d^c$. If all four
terms are present proton decay becomes very rapid. This problem is
circumvented by simply forbidding the last term, e.g.\ by using baryon
triality or a similar symmetry \cite{Dreiner:1997uz,Dreiner:2005rd}.
The remaining three terms  break lepton
number explicitly implying that a  combination of tree and loop
diagrams in these models can lead to realistic neutrino masses and
mixings.  From
the point of view of collider physics, there is an important
implication of LSP decay as at least some of its decay
properties are correlated to neutrino physics
 since the same couplings governing neutrino physics also
lead to visible decays of the LSP.

We will exemplify this by focusing on bilinear $R_P$ breaking, for discussion of
tri-linear $\rpv$ see for example
\cite{allanach:1999bf,Barbier:2004ez} and for the so-called
$\mu\nu$SSM see for example \cite{Bartl:2009an}. The absence of tri-linear terms
could be explained, for example, if bilinear R-parity breaking is the
effective low-energy limit of some spontaneous $\rpv$ model, see
below.

\subsection{Explicit bilinear R-parity violation}

The superpotential of the bilinear $\rpv$ model can be written as
\begin{equation} \label{spot}
{\cal W} =  W_{MSSM} + \epsilon_i \widehat L_i \widehat H_u .
\end{equation}
In addition, one must include bilinear $\rpv$ soft supersymmetry
breaking terms
\begin{equation} \label{vsoft}
V_{\rm soft} = \epsilon_i B_i {\tilde L}_i H_u + V_{\rm soft}^{MSSM}.
\end{equation}
These terms induce mixings between the MSSM Higgs
bosons and the left scalar neutrinos which in consequence also obtain
vevs $v_i$ once electro-weak symmetry
is broken.
Usually one trades the $B_i$ by the $v_i$ 
 using the corresponding tad-pole equations  which are a consequence
of  eq.~(\ref{vsoft}) as the connections to neutrino physics become more 
apparent.  

The effective
neutrino mass matrix at tree-level can then be cast into a very simple
form
\begin{equation}
m_{\nu,ij} = -
\frac{m_{\gamma}}{4 \mathrm{det}(M_{\chi^0})} \Lambda_i \Lambda_j
\label{eq:eff}
\end{equation}
The ``photino'' mass parameter is defined as $m_{\gamma} = g^2M_1
+g'^2 M_2$, $det(M_{\chi^0})$ is the determinant of the ($4,4$)
neutralino mass matrix and $\Lambda_{i} \equiv
\epsilon_{i}v_d+v_{i}\mu$ are the ``alignment
parameters''.

Due to the projective nature of eq.~(\ref{eq:eff}) the other two
neutrino masses are generated only at 1-loop order.  Generally the
most important contributions come from loops with sbottoms 
and staus \cite{Hirsch:2000ef,Diaz:2003as}. However, there exist also
parameter regions in the general $\rpv$ MSSM where the
sneutrino-anti-neutrino loop gives a sizeable contribution
\cite{Grossman:2000ex,Dedes:2006ni,Dedes:2007ef}.  
One finds that in order to explain the
observed neutrino mixing angles one requires certain relations among
the $\rpv$ parameters to be satisfied~\cite{Hirsch:2000ef}, e.g.\
the maximal atmospheric angles requires $\Lambda_{\mu}\simeq
\Lambda_{\tau}$.
\begin{figure}[t]
\begin{minipage}{0.47\textwidth}
\includegraphics[width=67mm,height=50mm]{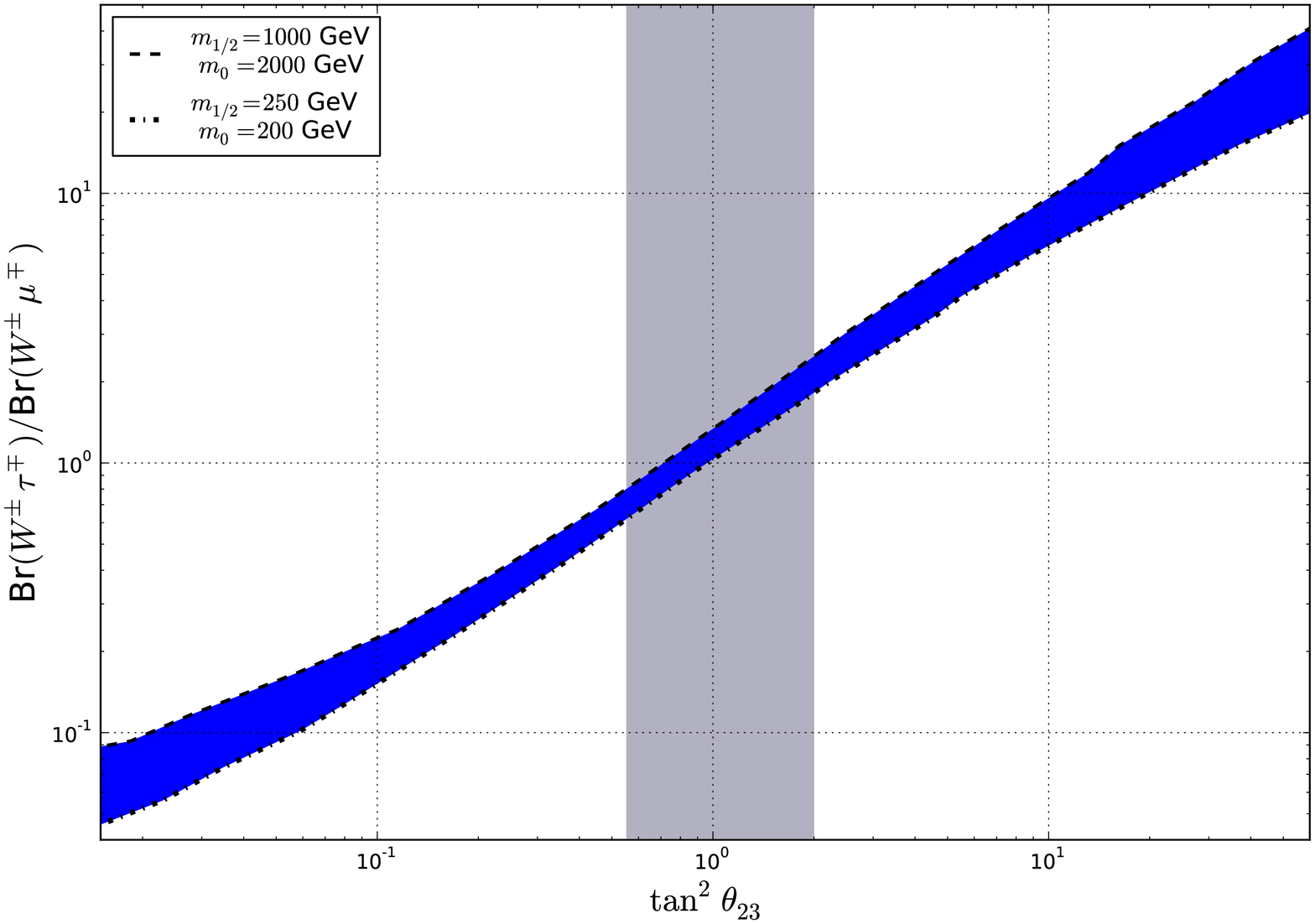}
\caption{Ratio of semi-leptonic branching ratios, BR$(\chi^0_1\to
\mu
  q'{\bar q})$ over Br$(\chi^0_1\to \tau q'{\bar q})$ as a function of
  the atmospheric neutrino angle calculated within bilinear $\rpv$
  SUSY, see ref.~\cite{Porod:2000hv}.}
\label{fig:ntrl}
\end{minipage}\hspace{0.04\textwidth}%
\begin{minipage}{0.47\textwidth}
\includegraphics[width=75mm,height=60mm]{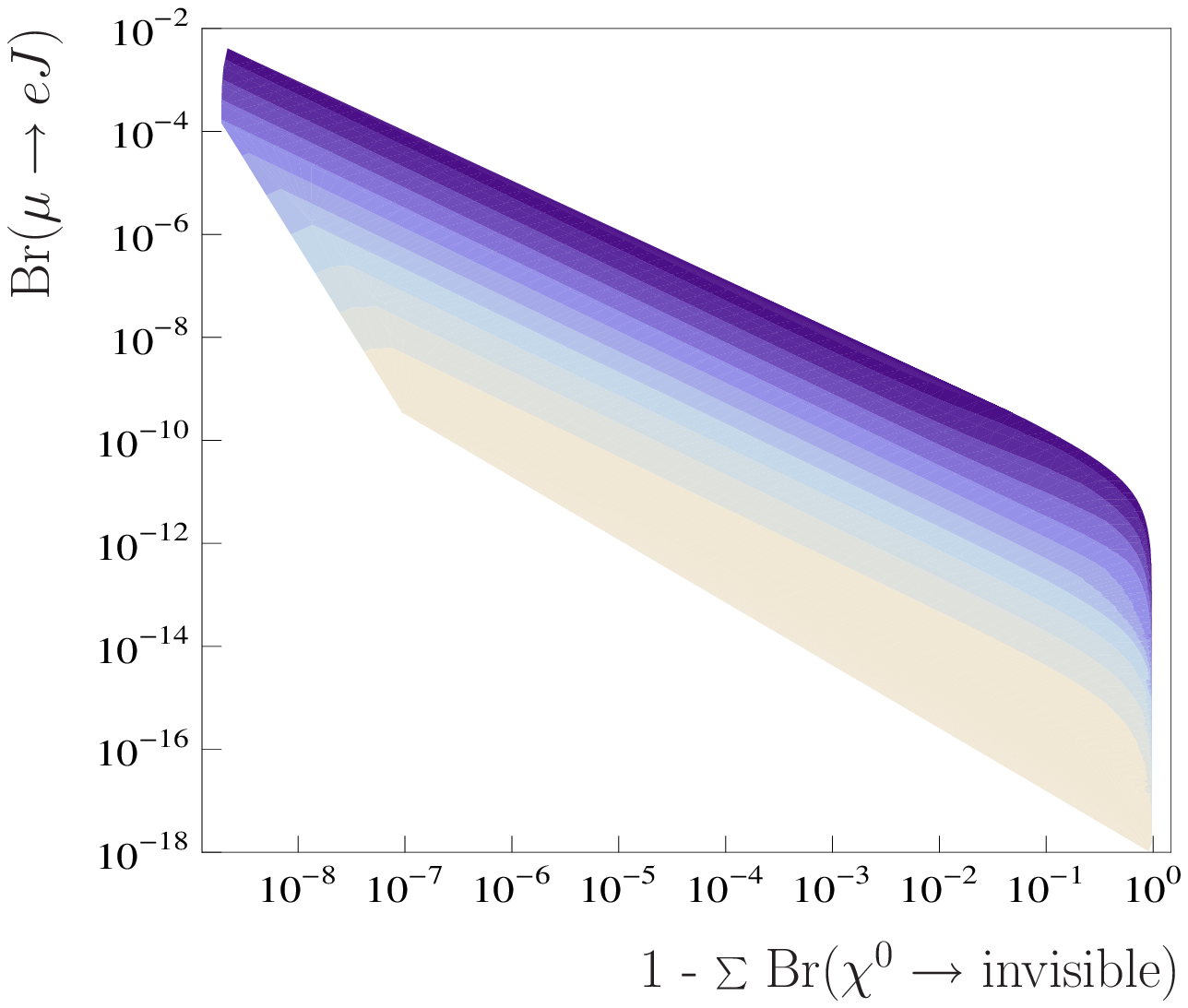}
\caption{Branching ratio BR($\mu\to eJ$) versus visible lightest
neutralino decay. $\mu\to eJ$ and $\chi^0_1\to J \nu$ are correlated,
see ref.\ \cite{Hirsch:2009ee}.}
\label{fig:VisMuEJ}
\end{minipage} 
\end{figure}

Once R-parity is broken the LSP decays. The decays of a neutralino LSP
have been studied in \cite{mukhopadhyaya:1998xj,Porod:2000hv}. Decay
lengths for the neutralino are approximately fixed once the neutrino
masses are fitted to experimental data. Typical lengths range from
tens of cm for very light neutralinos to sub-millimeter for
neutralinos of several hundred GeV \cite{Porod:2000hv}.  One of the
most exciting aspects of bilinear $\rpv$, however, is the fact that
once neutrino angles are fitted to the values
required~\cite{Schwetz:2008er} by the neutrino
oscillation data, the ratios of LSP decay branching ratios are fixed
and correlate with the observed neutrino mixing angles, as illustrated
for example in fig.~\ref{fig:ntrl}.  Measurements at the LHC should
allow to test this prediction, if signals of SUSY are found \cite{DeCampos:2010yu}.

Within $\rpv$ SUSY any supersymmetric particle can be the LSP.  It has been shown that
within bilinear $\rpv$ correlations between the measured neutrino
angles and ratios of LSP decays can be found for all LSP candidates
\cite{Hirsch:2002ys,Restrepo:2001me,Hirsch:2003fe}. Thus, it is 
possible to exclude the
minimal bilinear $\rpv$ model experimentally at the LHC.

\subsection{Spontaneous R-parity violation}

In spontaneous R-parity violation
models~\cite{Aulakh:1982yn,Ross:1984yg,santamaria:1989ic} R-parity
violation results from the minimization of the Higgs potential through
nonzero sneutrino vacuum expectation values. If lepton number is
ungauged this implies the existence of a
Nambu-Goldstone boson - the majoron. However, a doublet majoron is
ruled out  by LEP measurements of the Z
width~\cite{Amsler:2008zzb}. Hence, viable spontaneous R-parity
breaking models must be characterized by two types of sneutrino vevs,
those of right and left sneutrinos, singlets and doublets under
\321
respectively~\cite{masiero:1990uj,romao:1992vu}. These
obey the ``vev-seesaw'' relation $v_L v_R \sim Y_\nu m_W^2$ where
$Y_\nu$ is the small Yukawa coupling that governs the strength of
R--parity violation~\cite{masiero:1990uj,romao:1992vu}.

In this case the majoron is so weakly coupled that bounds from LEP and
astrophysics~\cite{Raffelt:1996wa} are easily satisfied.  For example,
the superpotential of \cite{masiero:1990uj} can be written as
\begin{eqnarray} %
W &=& \widehat H_u \widehat Q Y_u  \widehat U
          +  \widehat H_d \widehat Q Y_d\widehat D
          +  \widehat H_d \widehat L Y_e \widehat E 
          + \widehat H_u \widehat L Y_\nu \widehat \nu^c
          - h_0 \widehat H_d \widehat H_u \widehat\Phi
          + h \widehat\Phi \widehat\nu^c\widehat S +
          \frac{\lambda}{3!} \widehat\Phi^3 . 
\label{eq:Wsuppot}
\end{eqnarray}
The first three terms are the usual MSSM Yukawa terms. The terms
coupling the lepton doublets to $\widehat\nu^c$ fix lepton number.
The coupling of the field $\widehat\Phi$ with the Higgs doublets
generates an effective $\mu$-term a l\'a the Next to Minimal
Supersymmetric Standard Model (NMSSM)
\cite{Barbieri:1982eh,Nilles:1982dy,Chamseddine:1982jx}.  
Note, that $v_R = \langle \tilde \nu^c \rangle \ne 0$ 
generates effective bilinear terms $\epsilon_i
= Y_{\nu}^{i} v_{R} $ and that $v_R$, $v_S$ and $v_{L_i}$ violate
lepton number and R-parity spontaneously. 
Neutrino oscillation data enforce
$v_{L_i}^2 \ll v_R^2$ and $v_{L_i}^2 \ll v_d^2+v_u^2$ implying
 the majoron is mainly a singlet in this model \cite{Hirsch:2004rw,Hirsch:2005wd}.

The existence of the majoron affects the phenomenology at colliders mainly in
two ways: (i)  the lightest Higgs
can decay invisibly into two majorons \cite{Hirsch:2004rw,Hirsch:2005wd}. 
(ii) Also the decays of the
lightest neutralino are affected, since the new decay channel
$\chi^0_1\to J \nu$ is invisible at colliders. In ref.\
\cite{Hirsch:2006di,Hirsch:2008ur} it has been shown that this mode can be close
to 100\% for certain parameter combinations implying that a large
luminosity at the LHC might be required to detect R-parity violation. 
However,
as can be seen in fig.~\ref{fig:VisMuEJ} $chi^0_1\to J \nu$  is
correlated with the decay $\mu\to J e$, thus allowing to probe
for a complementary part of parameter space \cite{Hirsch:2009ee} with low
energy experiments.

Spontaneous $R$-parity violation can also be obtained by enlarging the
gauge group 
\cite{Kuchimanchi:1993jg,Huitu:1994zm,Huitu:1997qu,
FileviezPerez:2008sx,Everett:2009vy,Barger:2008wn,Ji:2008cq}. In this
case the majoron would be the longitudinal component of an additional
neutral heavy vector boson which can be produced in s-channel processes
at the LHC.

\subsection{LHC studies}

In the following we mainly focus on mSUGRA models which are augmented
by bilinear R-parity breaking parameters at the electroweak scale implying that
one has  eleven free parameters, namely $m_0$, $M_{1/2}$, $\tan\beta$,
sign$(\mu)$, $A_0$, $\epsilon_i$, and $\Lambda_i$.
In order to fit current neutrino oscillation data, the effective
strength of R-parity violation must be small.
This implies that supersymmetric particle spectra are expected to be
the same as in the conventional R-parity conserving model, and that processes
involving single production of SUSY states~\cite{nogueira:1990wz} are
negligible at the LHC and that
the only differences occur due to the decays of the LSP.

In these scenarios the lightest neutralino is the LSP and its
 main decay channels  are
$\tilde{\chi}^0_1 
\to \nu \ell^+ \ell^-$ with $\ell =e$, $\mu$ denoted by $\ell \ell$;
$\tilde{\chi}^0_1 
\to \nu \tau^+ \tau^-$, called $\tau \tau$;
$\tilde{\chi}^0_1 
\to \tau \nu  \ell$, called $\tau \ell$.
$\tilde{\chi}^0_1 
\to \nu q \bar{q}$ denoted $jj$;
$\tilde{\chi}^0_1 
\to \tau q^\prime \bar{q}$, called $\tau jj$;
$\tilde{\chi}^0_1 
\to \ell q^\prime \bar{q}$, called $\ell jj$;
$\tilde{\chi}^0_1 
\to \nu b \bar{b}$, which we denote by $bb$;
$\tilde{\chi}^0_1 
\to \nu b \bar{b}$, which we denote by $bb$;
$\tilde{\chi}^0_1 
\to \nu \nu \nu$.

\begin{figure}[t]
 \includegraphics[width=0.46\textwidth]{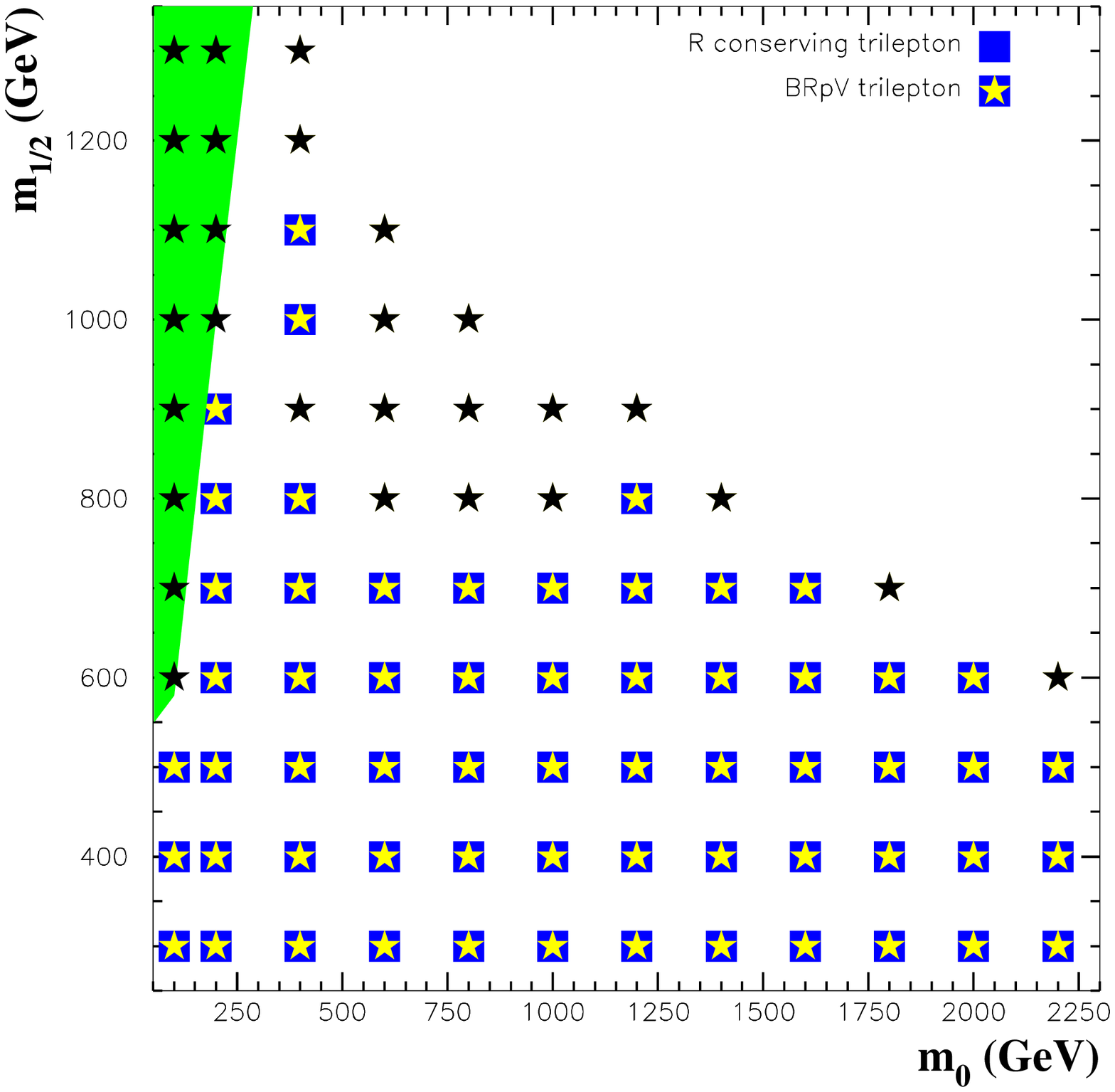}
 \hspace{0.06\textwidth}
 \includegraphics[width=0.46\textwidth]{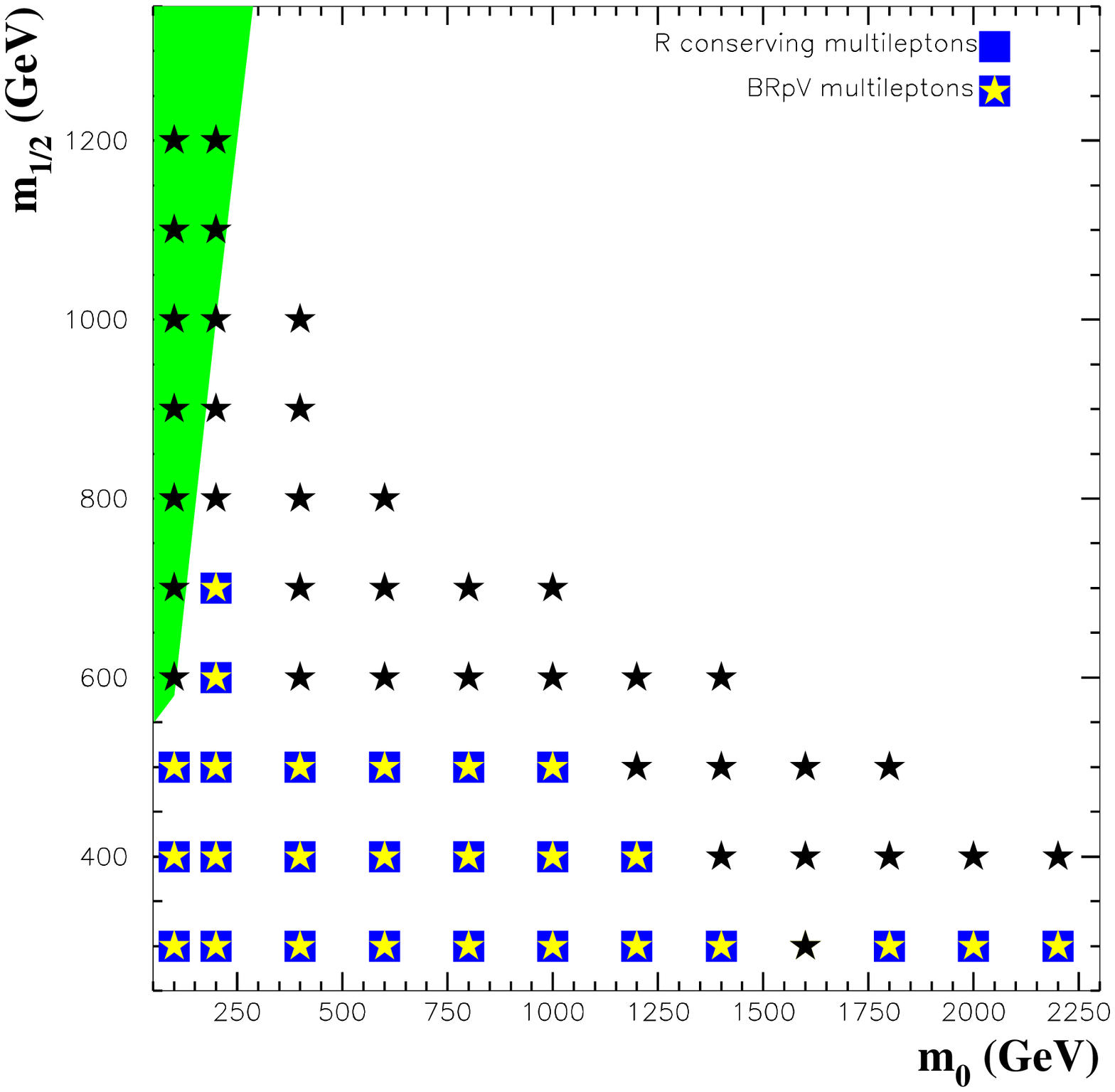}
  \caption{LHC discovery potential in the three lepton channel (top
    panel) and the multi-lepton one (bottom panel) for $A_0=-100$~GeV,
    $\tan\beta=10$, $\mu >0$ and an integrated luminosity of $100$
    fb$^{-1}$,  from
    ref.~\cite{deCampos:2007bn}. }
\label{fig:di:3lep100}
\end{figure}

In ref.~\cite{deCampos:2007bn} a comparison has been performed between
the reach of LHC for R-parity violating SUSY using the same cuts as
for R-parity conserving models \cite{Baer:2000bs}.
The main topologies are: Inclusive jets and missing transverse
momentum; zero lepton, jets and missing transverse momentum; one
lepton, jets and missing transverse momentum; opposite sign lepton
pair, jets and missing transverse momentum; same sign lepton pair,
jets and missing transverse momentum; tri-leptons, jets and missing
transverse momentum; multi-leptons, jets and missing transverse
momentum.  Due to the reduced missing energy the all-inclusive channel
will have a reduced reach in the parameter space. However, the decays
of the neutralino increase the multiplicities of the multi-lepton
channel. As an example we display in Fig.~\ref{fig:di:3lep100} the LHC
reach in the three-- and multi--lepton channels with/without R--parity
conservation for an integrated luminosity of 100 fb$^{-1}$. These
constitute the best standard channels for the discovery of bilinear
R-parity violation.

\begin{figure}[t]
\begin{minipage}{0.47\textwidth}
  \includegraphics[width=7.5cm]{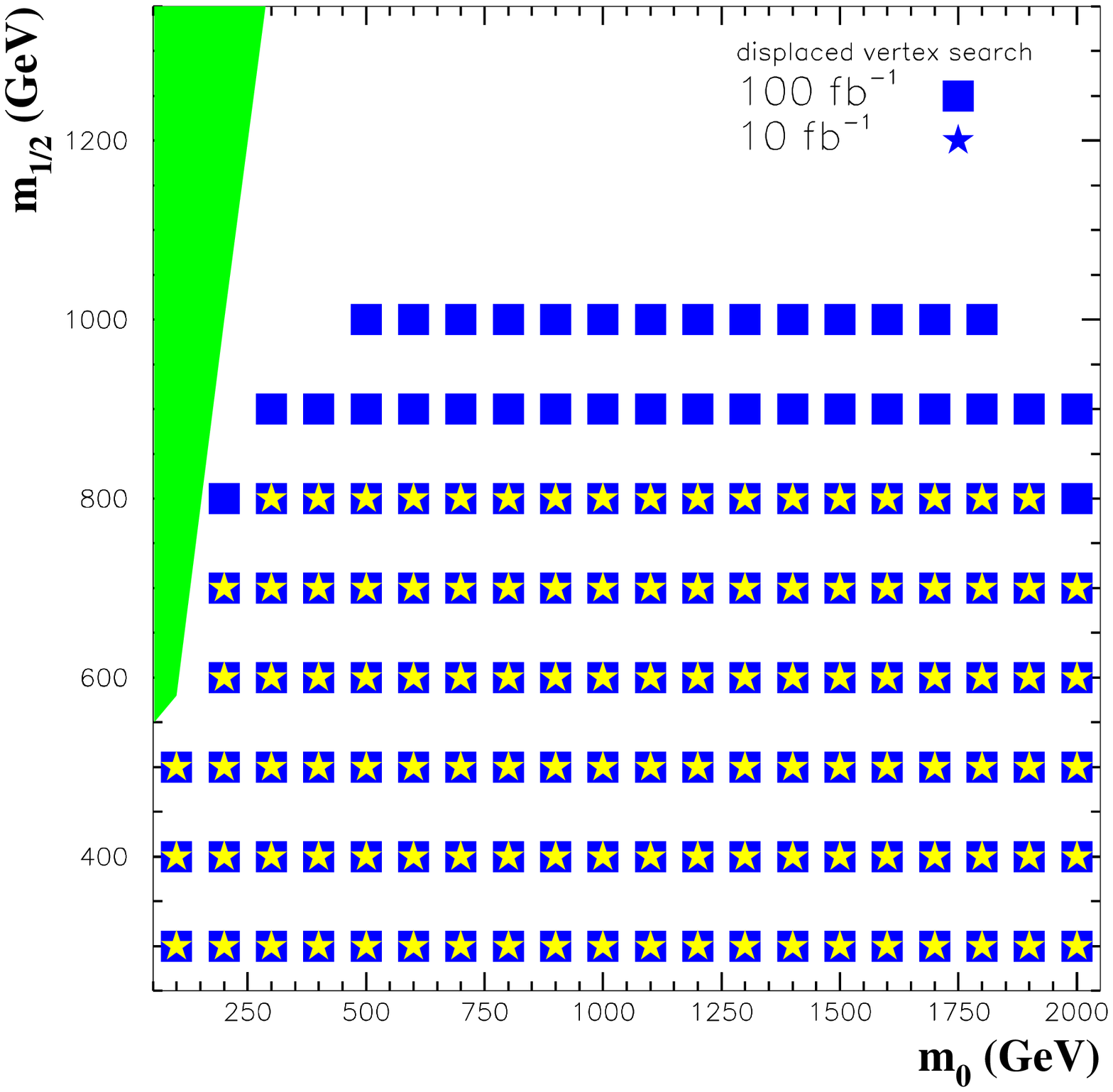}
  \caption{Discovery reach for displaced vertices channel in the
    $m_{0}$-$M_{1/2}$ plane for $\tan\beta = 10$, $\mu > 0$,
    $A_0=-100$ GeV. The stars (squares) stand for points where there
    are more than 5 displaced vertex signal events for an integrated
    luminosity of 10 (100) fb$^{-1}$.  The marked grey (green) area on
    the left upper corner is the region where the stau is the LSP,  from
    ref.~\cite{deCampos:2007bn}. }
 \label{fig:ms}
\end{minipage}\hspace{0.04\textwidth}%
\begin{minipage}{0.47\textwidth}
\includegraphics[width=7cm]{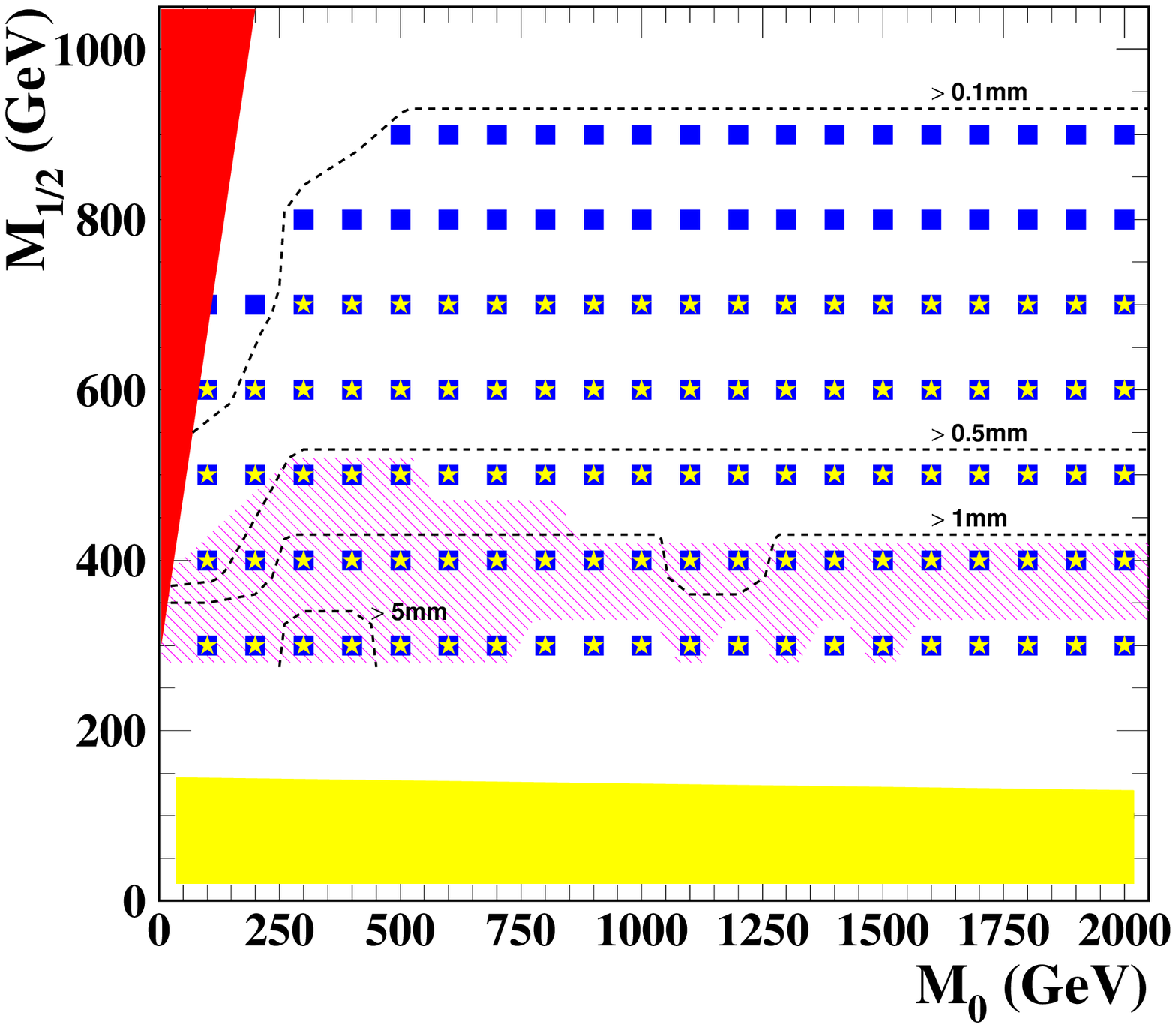}
  \caption{LHC reach for Higgs search  in the$m_{0}$-$M_{1/2}$  plane  for
    $\tan\beta = 10$, $A_0=-100$ GeV, and $\mu > 0$. The yellow stars
    (blue squares) show the reach for an integrated luminosity of
    10 (100) fb$^{-1}$ and the hatched region  the
    reach of LHCb for an integrated luminosity of 10
    fb$^{-1}$.  The (yellow) shaded region in the bottom is
    excluded by direct LEP searches, while the (red)
    upper--left area represents the region with a stau as
    LSP; the black lines delimit different regimes of LSP
    decay length,  from ref.\ \cite{deCampos:2008ic}.}
\label{fig:reach}
\end{minipage} 
\end{figure}

The sizable decay length of the neutralino is quite useful as
this topology has little, if any, background expected at the LHC. This
feature has been exploited in
ref.~\cite{deCampos:2007bn,deCampos:2008re}, where a comparison has
been performed between the reach of LHC for R-parity violating SUSY
including explicitly the displaced vertex topologies.
In figure~\ref{fig:ms} we present the displaced vertex reach in
the $m_{0}$-$M_{1/2}$ plane for $\tan\beta = 10$, $\mu > 0$,
    $A_0=-100$ GeV.  As one
can see form this figure, the LHC will be able to look for the
displaced vertex signal up to $m_{1/2} \sim $800 (1000) GeV
 for a large range of $m_0$
values and an integrated luminosity of 10 (100) fb$^{-1}$. Notice that
the reach in this channel is rather independent of $m_0$.
However, this signal  disappears in the region where
the stau is the LSP due to its rather short lifetime.

Next we discuss a tantalizing possibility, namely a double discovery
at the LHC: (i) find evidence for supersymmetry, and (ii) uncover the
Higgs boson.
There are regions in parameter space where
$\tilde{\chi}_1^0$ may have a sizeable branching ratio up to
22\% into the channel $\nu h^0$ where $h^0$ is the lightest Higgs boson
\cite{deCampos:2008ic}.
This would lead to displaced vertices containing two b-jets as a
characteristic signature for Higgs production at the
LHC~\cite{deCampos:2008ic}.
The displaced vertex signal implies that also LHCb will have good
sensitivity for such scenarios in particular in case of final states
containing muons such as $\tilde{\chi}^0_1 \to \nu \mu^+ \mu^-$.%
Figure~\ref{fig:reach} demonstrates that the ATLAS and CMS experiments
will be able to look for the signal up to $M_{1/2} \sim 700$ $(900)$
GeV for a LHC integrated luminosity of 10 (100) fb$^{-1}$.
The hatched region in Fig.~\ref{fig:reach} indicates the LHCb reach
for 10 fb$^{-1}$. Due to the strong cut on the pseudo--rapidity
required by this experiment the reach for 2 fb$^{-1}$ is severely
depleted and only a small region of the parameter space is covered.

\subsection{Trilinear R-parity breaking}

Bilinear $R$-parity violation is essentially equivalent to tri-linear
$R$-parity breaking with the superpotential
\begin{eqnarray}
W_{tri} = {\textstyle \frac{1}{2}}
 \lambda_{ijk}  \widehat L_i  \widehat L_j\widehat E_k
 + \lambda_{ijk}' \widehat L_i\widehat Q_j\widehat D_k
\end{eqnarray}
where the tri-linear couplings have the following structures
\begin{eqnarray}
\lambda_{ijk} \simeq \frac{\epsilon_i}{\mu} Y^{jk}_e \,\,,\,\,
\lambda_{ijk}' \simeq \frac{\epsilon_i}{\mu} Y^{jk}_d
\end{eqnarray}
Obviously the phenomenology will be very similar if tri-linear
$R$-parity violation is close to this structure. In the case of
significant deviations from this structure is realized, one
gets new interesting signatures. For example there exists light
stau LSP scenarios where $\tilde \tau_1$ decays dominantly  via
4-body decays such as $ \tilde \tau_1 \to \tau^- \mu^- u \bar{d}$
with long lifetimes leading to displaced vertices \cite{Dreiner:2008rv}.
Another interesting signals are the resonant production of sleptons
as discussed in \cite{Dreiner:2000vf,Moreau:2000bs} or
associated production of single sleptons with $t$-quarks
\cite{Bernhardt:2008mz}.

An interesting question is to which extent one can measure deviations
from the hierarchical structure above, e.g.~the coupling $\lambda_{211}'$
can still be of order $0.1$. It has been shown in \cite{Choudhury:2002av}
that in such a case one LHC will be able to measure such couplings of such
a strength with an accuracy of about 10\%.

\section{Summary}

We have discussed aspects of various extensions of the MSSM taking the
explanation of neutrino data as guide-line. As a first model the case
has been considered where one adds a Dirac mass term for the neutrinos
with tiny Yukawa couplings. In this case one obtains the usual MSSM
phenomenology except for the case where the right-sneutrino is the LSP
because in this case the NLSP will be very long-lived and one gets as
signature quasi-stable particles at the LHC.

In case of seesaw models one finds that in case of type II and type
III models the spectrum can be quite different when compared to the
usual mSUGRA scenarios as a consequence of additional charged
particles between the seesaw scale(s) and the GUT scale. These
differences are the larger the lower the seesaw scale and affect in
particular the masses of the sfermions.  One can study four ratios at
the LHC which can give information on the GUT scale if one assumes
universal boundary conditions for the soft SUSY breaking parameters at
the GUT scale. Moreover, there are parameter regions in all seesaw
models where lepton flavour violating signals can be found in SUSY
cascade decays. However, for this in general a rather high luminosity
in the order of 100 $fb^{-1}$ or higher is required.

Supersymmetric models also offer the possibility to explain neutrino
data via R-parity violation. Here we have mainly focused on models
where R-parity is broken by bilinear terms as this is sufficient to
explain neutrino data and to explore the relevant LHC
phenomenology. Moreover, this class of models can be obtained as
effective model in case of spontaneous R-parity breaking. Bilinear
R-parity breaking implies correlations between neutrino data and
decays properties of the LSP, e.g.~in case of an neutralino LSP the
ratio $BR(\tilde{\chi}^0_1 \to W^- \tau^+)/BR(\tilde{\chi}^0_1 \to W^- \mu^+)
\simeq
\tan^2\theta_{atm}$ where $\theta_{atm}$ is the mixing angle related to the
the atmospheric neutrino sector.
In addition one finds very often that the LSP life time is
sufficiently small to produce a displaced vertex in a typical collider
experiments.

\section*{Acknowledgments}
This work has been supported by  the DFG, project number PO-1337/1-1,
 the Alexander von Humboldt Foundation and
the Spanish grant FPA2008-00319/FPA.


\end{document}